# A Low-Power 9-bit Pipelined CMOS ADC with Amplifier and Comparator Sharing Technique

Yuri Bocharov, Vladimir Butuzov, Dmitry Osipov
Department of Micro and Nanoelectronics
National Research Nuclear University MEPHI
Moscow, Russian Federation
DLOsipov@mephi.ru

*Abstract*— This paper describes a pipelined analog-to-digital converter (ADC) employing a power and area efficient architecture. The adjacent stages of a pipeline share operational amplifiers. In order to keep accuracy of the amplifiers in the first stages, they use a partially sharing technique. The feature of the proposed scheme is that it also shares the comparators. The capacitors of the first stages of a pipeline are scaled down along a pipeline for a further reducing the chip area and its power consumption. A 9-bit 20-MSamples/s ADC, intended for use in multi-channel mixed-signal chips, has been fabricated via Europractice in a 180-nm CMOS process from UMC. The prototype ADC shows a spurious-free dynamic range of 58.5 dB at a sample rate of 20 MSamples/s, when a 400 kHz input signal with a swing of 1 dB below full scale is applied. The effective number of bits is 8.0 at the same conditions. ADC occupies an active area of 0.4 mm$^2$ and dissipates 8.6 mW at a 1.8 V supply.

## I. INTRODUCTION

Power and area efficiency is a common problem in the design of analog-to-digital converters (ADC). Today power lowering has moved to the forefront of design challenges. There are some applications where the achievement of very low power consumption of ADC is critical. In particular, this is the instrumentation for modern physics experiments. For instance, the front-end electronics of the Silicon Tracking System of the upcoming CBM experiment in GSI/FAIR at Darmstadt contains more than one million readout channels [1] and [2]. The implementation of so large-scale systems requires using highly integrated mixed-signal application-specific integrated circuits (ASIC), which contain multiple ADCs. In some other ADC applications, such as battery-powered measuring devices and communication units, very low power consumption at high and moderate sample rates is also mandatory.

In these applications pipeline architecture has been widely adopted because it guarantees high speed with reasonable requirements for the resolution of comparators, an acceptable power consumption and small area. A number of methods have been proposed for reducing power dissipation and silicon area of pipelined ADC, including the technique of sharing amplifiers by adjacent stages of a pipeline [3]–[10]. This research focuses on a study of methods of further reducing the power dissipation of the pipelined ADC.

The design of a 9-bit 20-MSamples/s ADC utilizes the proposed methods. The prototype ADC has been fabricated in a 180-nm CMOS process. To minimize power consumption the ADC uses a state-of-the-art amplifier sharing technique as well as the proposed comparator sharing technique. The paper considers the advantages and drawbacks of this approach.

## II. CIRCUIT OVERVIEW

The ADC has a fully differential pipelined architecture. Fig. 1 shows its block-diagram. The signal conversion path is shown single-ended for simplicity. The auxiliary blocks and peripherals such as band-gap voltage reference, on-chip reference voltage buffers, clock generator and digital interface are not shown.

The pipeline includes seven 1.5-bit stages and a 2-bit flash ADC. All 1.5-bit stages except the first and the second are identical. They are grouped, as shown in Fig. 1. Some elements of the two adjacent stages belong to only one stage, and some of their elements are common. This makes it possible to reduce dramatically the power consumption as well as the chip area. Sharing of operational amplifiers, which are the key elements of the switched-capacitor multiplying digital-to-analog converter (MDAC), is a commonly used technique. The feature of the proposed solution is that the amplifiers of the first stages are shared only partially and the comparators are shared too.

## III. POWER REDUCTION TECHNIQUES

The possibility of sharing some elements of the adjacent stages in the pipelined ADCs is because at any time the states of the odd-numbered stages are different from states of even-numbered stages. In one-half of a clock cycle, the odd stages are in a sampling phase and the even stages are in the amplification and residue estimation mode.

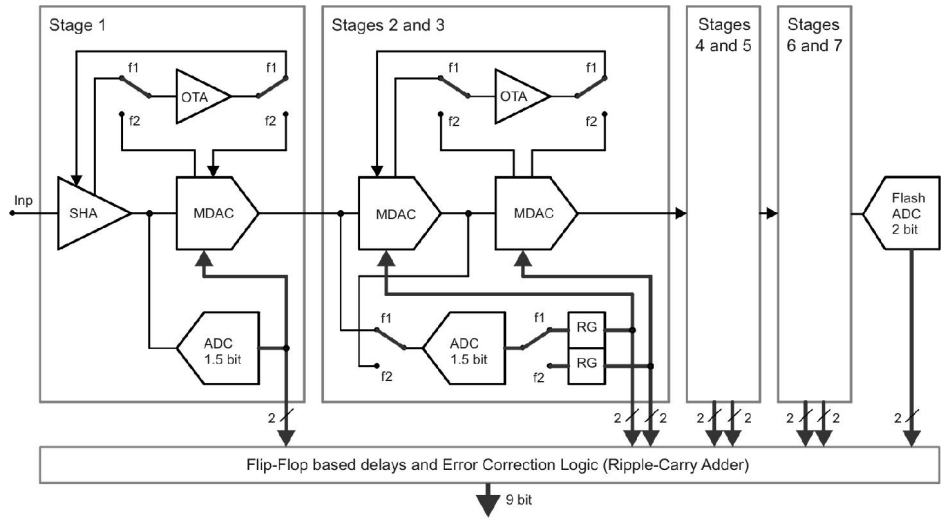

Figure 1. Proposed ADC architecture.

During the other half of a cycle, they go to an opposite state. The switched-capacitor circuits can afford to share the amplifier between adjacent stages because they require the amplifier only in the amplification phase and do not require in the sampling phase. During the sampling phase operational amplifiers are either in the offset correction mode or not used at all. If offset correction is not required, the amplifiers can be switched alternately between adjacent stages to be connected to the stage, which is in a residue estimation mode. As the adjacent stages are always in opposite states, amplifiers and comparators can be allocated to the common shared resource. This allows halving the number of amplifiers.

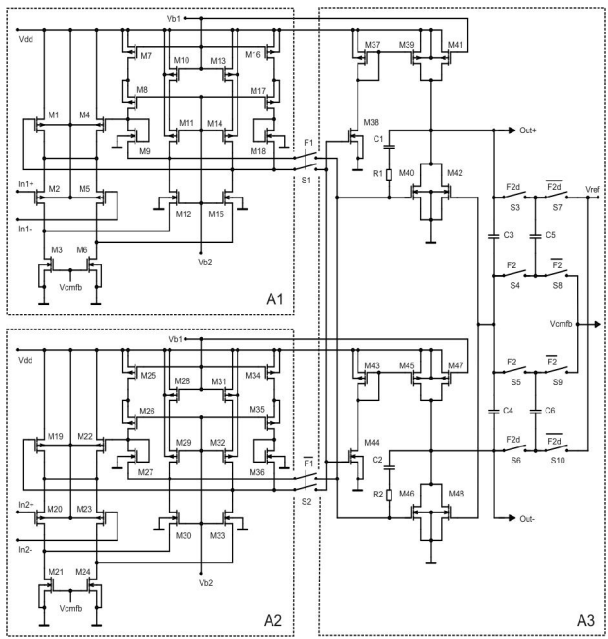

Figure 2. Schematic diagram of the partially shared fully differential operational amplifier.

In the proposed ADC the amplifiers, which are shared by the input sample-and-hold amplifier (SHA) and the first MDAC, as well as the amplifiers of the second and third stages are shared partially. This compensates the amplifier-offset-voltage and suppresses a memory effect in capacitors.

Fig. 2 illustrates the partially shared amplifier. It contains two identical not shared preamplifiers A1 and A2, connected through the switches S1 and S2 to a shared block A3, which includes an output amplifier, a switched-capacitor common-mode feedback circuit and a bias circuit. In a sampling phase, A1 and A2 are in the offset compensation mode while A3 is connected to the preamplifier of the adjacent stage. Amplifiers in the stages from fourth to seventh are completely shared.

Comparators are also shared between adjacent stages. Fig. 3 illustrates a block-diagram of a circuitry that contains a differential comparator shared by the stages numbered as i and i+1. The comparator schematic diagram is shown in Fig. 4. It is a regenerative latch with a built-in preamplifier, clocked by non-overlapped timing sequences shown in Fig. 5.

The stages can share the comparator because after latching of its output in flip-flop registers at the beginning of a sampling phase, during the rest of this phase, comparator is not active and therefore it can be used in the adjacent stage, which is in the residue estimation mode.

The drawback of this solution is that a latch signal should have a frequency twice the ADC sampling rate. It requires an extra low-power frequency-doubling block. Fig. 6 shows a schematics and a timing diagram of the developed frequency doubler. If the input signal is a 50% duty cycle clock and the capacitors C1 and C2 are tolerably matched, the frequency multiplication factor keeps its value close to two in a wide range of the input frequencies.

The proposed technique allowed to reduce the number of comparators from 17 to 11 as well as to reduce their total power consumption and occupied area, adjusted for the effect of the frequency doubler, by more then 30%.

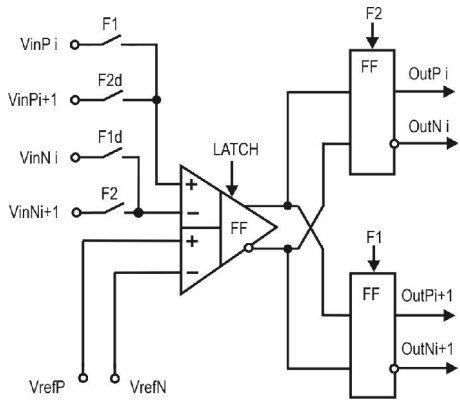

Figure 3. Block-diagram of the shared comparator.

The sampling capacitors of the first stages of the chain are scaled down along a pipeline. The SHA and MDAC of the first stage use the capacitors of 0.8 pF, while the MDAC of the fourth and further stages use the capacitors of 0.25 pF. This results in a lower area and relaxed requirements for operational amplifiers in the stages from fourth to seventh. They utilize the amplifiers with minimal operating currents and consequently with a minimal power consumption [11].

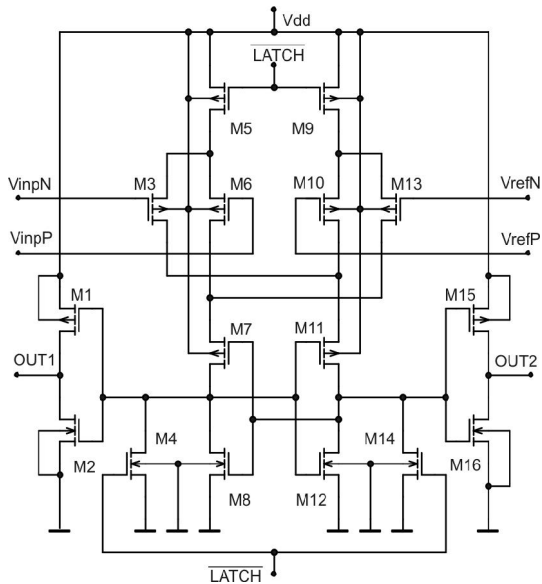

Figure 4. Schematic diagram of the comparator.

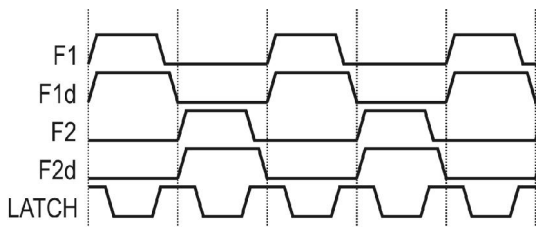

Figure 5. ADC clocking.

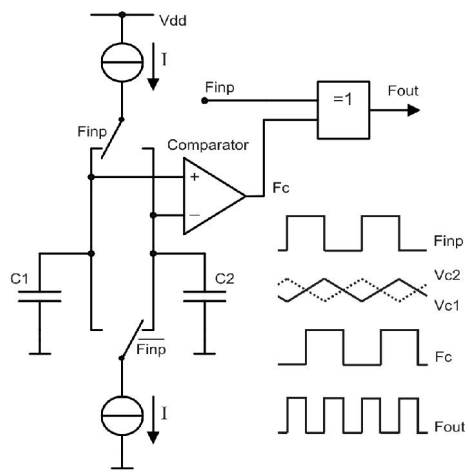

Figure 6. Schematics and timing-diagram of the frequency doubler.

## IV. IMPLEMENTATION

The ADC was designed in the Cadence IC6.14 Virtuoso environment as an intellectual property (IP) block of the future mixed-signal multi-channel system-on-chip (SoC). Its first application is expected to be in a silicon tracker station of the CBM experiment.

A prototype ADC has been fabricated via Europractice at the UMC foundry in 180-nm CMOS process focused on mixed-mode and radio frequency circuits. It is a single-poly, 6-metal layers (1P6M) process with metal-insulator-metal (MIM) capacitors. The ADC occupies an active silicon area of 0.4 mm$^2$.

At a sampling frequency of 20 MHz the ADC has a power consumption up to 14 mW. The core dissipates no more than 8.6 mW. The rest is the consumption of the reference voltage source and reference voltage buffers, which are common blocks for multiple channels.

Fig. 7 illustrates the FFT spectrum of ADC output when a sinusoidal input signal with the swing of 1 dB below full scale is applied. Coherent sampling was used to suppress a spectral leakage. Fig. 8 and Fig. 9 show the ADC layout and a die photograph.

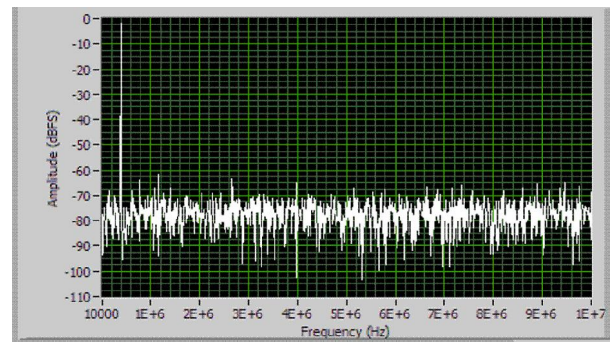

Figure 7. FFT spectrum at input frequency of 400.39 kHz and sampling frequency of 20 MHz.

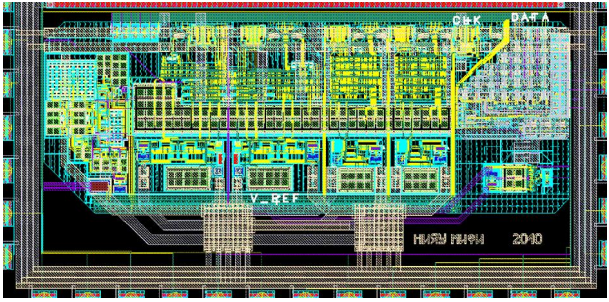

Figure 8. ADC layout.

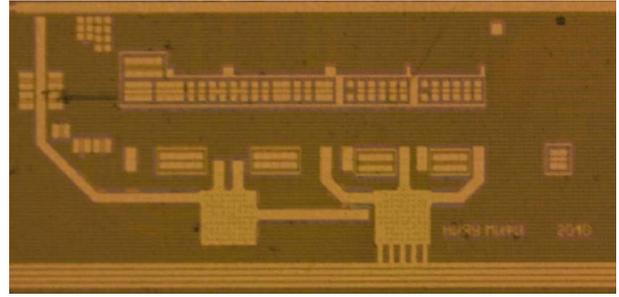

Figure 9. Die photograph.

## V. CONCLUSION

This paper describes a low power pipelined ADC for a multi-channel mixed-signal readout ASIC which is intended for use in CBM silicon tracking system at GSI/FAIR facility [12]. By sharing amplifiers and comparators along a pipeline, the proposed 9-bit ADC uses only 4 amplifiers and 11 comparators instead of 8 amplifiers and 17 comparators with the conventional pipeline architecture. More then 30% of the total power consumption of comparators and their occupied area is reduced. A frequency doubler for clocking the shared comparators was proposed. The partial sharing of amplifiers in the first stages of a pipeline enabled to keep their accuracy. The prototype ADC has been fabricated in a 180-nm CMOS process. The device reaches a spurious free dynamic range of 58.5 dB at the sample rate of 20 MHz when the input signal of 400 kHz with the swing of 1 dB below full scale is applied. The effective number of bits (ENOB) is 8.0 at the same conditions. The core power consumption is 8.6 mW. It occupies of an active area of 0.4 mm$^2$. Table I summarized the ADC performance.

TABLE I. ADC PERFORMANCE SUMMARY

| Parameter | Unit | Value |
|---|---|---|
| Resolution | bit | 9 |
| Sampling Rate | MSamples/s | 20 |
| Effective Number of Bits (ENOB) at 400 kHz input | bit | 8.0 |
| Spurious Free Dynamic Range (SFDR) at 400 kHz input | dB | 58.5 |
| Full Scale (differential) | $V_{p-p}$ | 2 |
| Supply Voltage | V | 1.8 |
| Core power consumption | mW | 8.6 |
| Total power including on-chip reference voltage buffers | mW | 14 |
| On-chip reference source voltage | V | 1.23 |
| Active area | mm$^2$ | 0.4 |
| Technology | | Mixed Mode 180-nm CMOS process with MIM capacitors |